\documentclass[prl,twocolumn,showkeys,superscriptaddress,nofootinbib]{revtex4-1}
\usepackage{amsmath}
\usepackage{amsfonts}

\usepackage{xr}
\usepackage{xr-hyper}
\usepackage[colorlinks,urlcolor=blue,linkcolor=blue]{hyperref} 
\usepackage{graphicx}

\usepackage[caption=false]{subfig}

\begin{document}
\title{Emergence of Metachronal Waves in Active Microtubule Arrays} 
\author{Stephen E Martin}
\affiliation{Department of Physics, University of California, Santa Cruz, CA 95064, USA}
\author{Matthew E Brunner}
\affiliation{Voltaic Inc. 2150 Shattuck Ave, \#704 Berkeley, CA 94704}
\author{Joshua M Deutsch}
\email{josh@ucsc.edu}
\affiliation{Department of Physics, University of California, Santa Cruz, CA 95064, USA}
\date{\today}
\begin{abstract}
\noindent
The physical mechanism behind the spontaneous formation of metachronal
waves in microtubule arrays in a low Reynolds number fluid has been
of interest for the past several years, yet is still not well
understood. We present a model implementing the hydrodynamic coupling
hypothesis from first principles, and use this model to simulate
kinesin-driven microtubule arrays and observe their emergent behavior.
The results of simulations are compared to known experimental
observations by Sanchez et al.\cite{Sanchez2011,sanchez2013engineering}. By varying
parameters, we determine regimes in which the metachronal wave
phenomenon emerges, and categorize other types of possible microtubule
motion outside these regimes.

\end{abstract}

\maketitle


Metachronal waves refer to the synchronization of thin, flexible
appendages that result in large-scale wavelike formations. These
appear in biological systems at the macroscopic scale (e.g. the
motion of millipede legs) and at the microscopic scale (e.g. cilia
in air pathways). On the microscopic level, metachronal waves are
essential components of several critical biological processes, from
motility in microorganisms to mucus clearance in human bronchial
tubes \cite{Afzelius2004,Okada2005}. If cilia are unable to effectively
move and synchronize, the results are often severe -- especially
if the disorder is genetic \cite{Afzelius2004}. Research into
physical explanations for cilia beating~\cite{brokaw1975molecular}, and of spontaneous metachronal behavior in cilia
is ongoing and still not well understood~\cite{camalet1999self,lindemann2010flagellar}, although many have suggested
that this phenomenon can be explained from hydrodynamic coupling
between cilia \cite{Sleigh1969,Sleigh1974,Gheber1989,Gueron1997}.

Recently, in some remarkable experiments, Sanchez et al. demonstrated metachronal wave behavior in
an \textit{in vitro} system\cite{Sanchez2011,sanchez2013engineering}.  
Microtubules (MTs) aggregated into bundles of length $10-100 \mu\mathrm{m}$ due to the addition of
polyethylene glycol~\cite{needleman2004synchrotron}.
Many of these bundles attached at one end
to a fixed boundary forming dense arrays. When exposed to a solution
containing clusters of kinesin and ATP, sustained metachronal wave
behavior between MT bundles (similar to that displayed by cilia and
flagella) was observed. MT bundles were constrained to move between
two glass slides. It is surprising that a system with such few ingredients
could develop complex behavior that so closely
resembles biological systems, which are made up from a much
more complicated machinery. Proteomic analysis indicate that eukaryotic cilia are
composed of many hundreds of proteins~\cite{pazour2005proteomic}.

Some important details of this
\textit{in vitro} system are still unclear, most notably whether the
MTs in this experiment are unipolar or of mixed polarity. 
Opposite polarity MTs will move past each other, causing separation
into unipolar bundles~\cite{kruse2000actively,liverpool2003instabilities}. 
We present analytical and numerical arguments for unipolarity in 
S-IV.
The surprising mechanism for the motion of unipolar bundles described here,
has not previously been given~\cite{Sanchez2011,sanchez2013engineering}, 
and we believe that the agreement between our model and
experiments provides further evidence to support our proposed explanation. 

The general mechanism proposed is quite similar
to the model used to describe and simulate cytoplasmic
streaming in \textit{Drosophila} oocytes, and the fact that it can
be adapted as such is in many ways a testament to its predictive
power. A fair amount of attention has been paid in recent years to
the understanding of how metachronal waves form in such arrays
\cite{Lagomarsino2003,guirao2007spontaneous,Elgeti2013,Niedermayer2008}. However, such
models often rely on assumptions about individual MT (or cilia)
beat patterns and/or on phenomenology. The model we propose makes
no such assumptions (beyond some minor simplifications), relying
on first-principles fluid mechanics calculations. This is important,
as it is not clear why one would want additionally to impose oscillatory behavior
on individual MTs given the lack of a well defined internal structure.

\begin{figure}
\includegraphics[width = 1.2in]{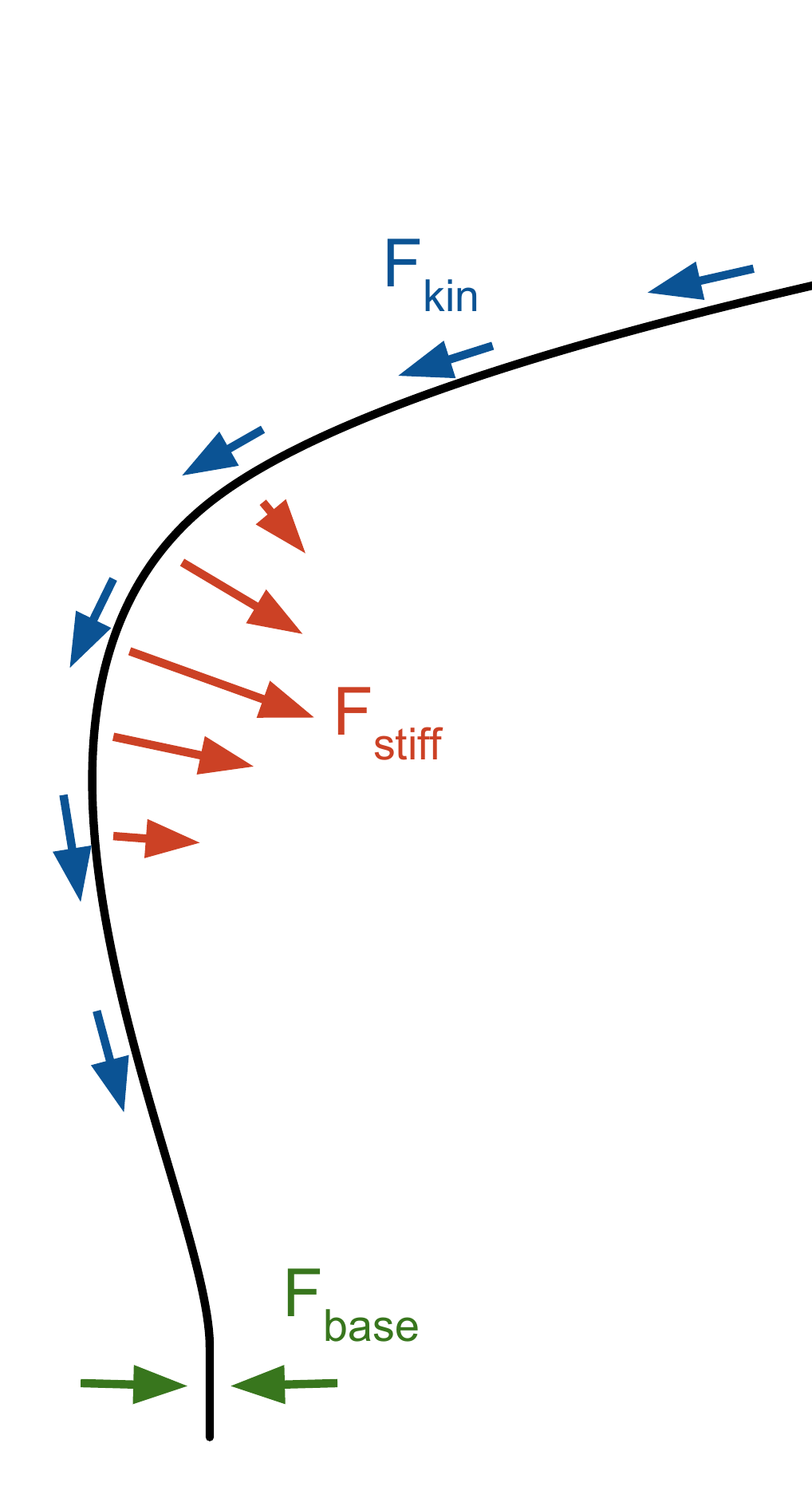}
\caption{Conceptual illustration of forces acting on a single polymer
that are not due to hydrodynamic interactions. The blue vectors
indicate the buckling forces due to kinesin walkers (tangent to
polymer), the red vectors show the direction and relative magnitude
of stiffness forces (in the direction of $d^4\mathbf{r}/ds^4$), and
the green arrows indicate a restorative force keeping the base of
the polymer approximately perpendicular to the binding
surface.
\label{fig:Mtforces}
}
\end{figure}

We now present a model for the simulation of the Sanchez
et al. system. A similar method has been used successfully to
simulate cytoplasmic streaming in \textit{Drosophila}
oocytes\cite{Monteith2016}, and is based on theoretical
work completed several decades ago regarding the calculation of
Stokes flows created by a point force (stokeslet) near no-slip
boundaries\cite{Blake1971,Liron1976}. A conceptual explanation of
this mechanism is given below, and further details regarding theory
and implementation are given in supplementary materials 
S-II and S-III.

An illustration of how MT bundles are simulated is given in Fig.
\ref{fig:Mtforces}. Each MT bundle is modeled as a chain of monomers
(i.e. polymer) which are held an approximately fixed distance from
one another by a spring force. The base of each polymer is anchored
to a single point, and the polymer at the base is kept roughly
perpendicular to the anchoring surface. Let the polymer be described
by the curve $\mathbf{r}(s)$, where $\mathbf{r}(0)$ is the location
of the polymer base, and $s$ is the arc length. We give the polymer a stiffness by implementing
an energetic cost of bending proportional to curvature squared, which
implies a local force at $s$ proportional to  $d^4\mathbf{r}/ds^4$. Additionally, monomers feel a ``buckling"
force due to the drag from the walking kinesin $F_{kin} = - f_k d\mathbf{r}/ds$, which is parallel to the polymer
and toward the polymer base. $f_k$ will depend linearly on the speed of the kinesin
and the solvent viscosity. This force continually adds energy to
the system (making it active), and has been shown to be a good
representation of the average drag force due to kinesin walking
along the microtubule away from the polymer base~\cite{Monteith2016}.

This kind of model for a single chain was first employed to understand glide assay dynamics
in two dimensions~\cite{bourdieu1995spiral}. In three dimensions,
periodic waves develop whose dynamics have been analyzed in detail~\cite{Monteith2016}, and
related theoretical work has recently also been performed~\cite{de2017spontaneous}. However,
scaling can be used to get the relevant length and timescales~\cite{bourdieu1995spiral}.
The average radius of curvature depends on the strength of the buckling force $f_k$,
and the elastic constant of a filament characterizing its stiffness $k_{stiff}$. The radius of curvature over
quite a wide range of parameters can be shown to be
$R = (k_{stiff}/(\beta f_k))^{1/3}$, where $\beta \approx 0.05$. Likewise, the angular frequency is
$\omega = f_k/(\nu R)$, where $\nu$ is the hydrodynamic drag coefficient per unit length.
Although there is a fairly large experimental uncertainty in parameters
used to model a Drosophila oocyte, this model finds quite good agreement with the experimental
time and length scales. $R$ was predicted to be $25-54\mu$m, close to the $16.3\pm 2.2 \mu$m observed.
Likewise, the time scale was predicted to be $203-1094$s, which is in the observed range of $370\pm 42$s.
It is interesting that the length and time scales observed by Sanchez et al. are also quite close to
these numbers, and that the frequency of biological cilia beating is often three orders of magnitude higher than this.

Polymers also feel hydrodynamic forces.
As the force from the kinesin causes the polymers to buckle, we
begin to see complex motion. Each monomer acts as a point force
(stokeslet) in the surrounding fluid. This force, that a monomer exerts
on the fluid, is simply the sum of all of the other forces on the
monomer: because the Reynolds number is nearly zero, there are no
inertial terms, meaning the force is transferred perfectly from the
monomer to the fluid. As this is a Stokes flow, the flow contributions
from all stokeslets add linearly, and we can (in principle) calculate
the flow everywhere. However, we only need to calculate the fluid
velocity at points with monomers. 
Therefore, the evolution can be calculated via a pairwise sum over
all monomers (see the supplemental materials 
S-II.

We also assume all polymer motion is two dimensional with a constant value of $z$, which
is physically sensible when considering the geometry of the Sanchez et al.
experiments. In this experiment, MT bundles were observed between
glass slides, with a height $H$, of approximately $10 \mu$m, creating a narrow channel for which fluid can flow.
For this reason, we adopt a two dimensional geometry. In addition,
the no-slip boundaries
of the plates have a large impact on the hydrodynamic forces between
monomers\cite{Blake1971,Liron1976}, which we give explicitly in
the supplemental materials 
S-II.
Other close-range contact forces were
also used (repulsion from anchoring surface, monomer-monomer
repulsion), and these are explained in the supplementary materials 
S-III.

We can now address at the qualitative level the mechanism by which
we propose the metachronal waves observed by Sanchez et al. form.
As kinesin walk away from the polymer bases, the polymers will tend
to buckle. If a polymer is isolated, this buckling will lead to
corkscrew motion or periodic
waves\cite{Monteith2016}. When placed in an array,
however, nearby polymers will exert hydrodynamic forces on one
another that tend to synchronize their motion. If these hydrodynamic
forces are sufficiently strong, this can cause a transition from
disordered motion to aligned MTs and correlated motion. 

Despite the fact that this model was developed to explain and
simulate cytoplasmic streaming, its mechanism 
can be easily adapted for related biological phenomena. Indeed,
when the conditions of the Sanchez et al. experiment are simulated
in the same way, we observe metachronal waves.
It is not clear if this is formally a transition or a more continuous
crossover effect, but the results found make strong predictions
that should be testable experimentally.
In the following, we present
the results of these simulations and discuss the required conditions
for metachronal wave formation.

\begin{figure}
\centering
\subfloat[][]{
\includegraphics[width =\columnwidth]{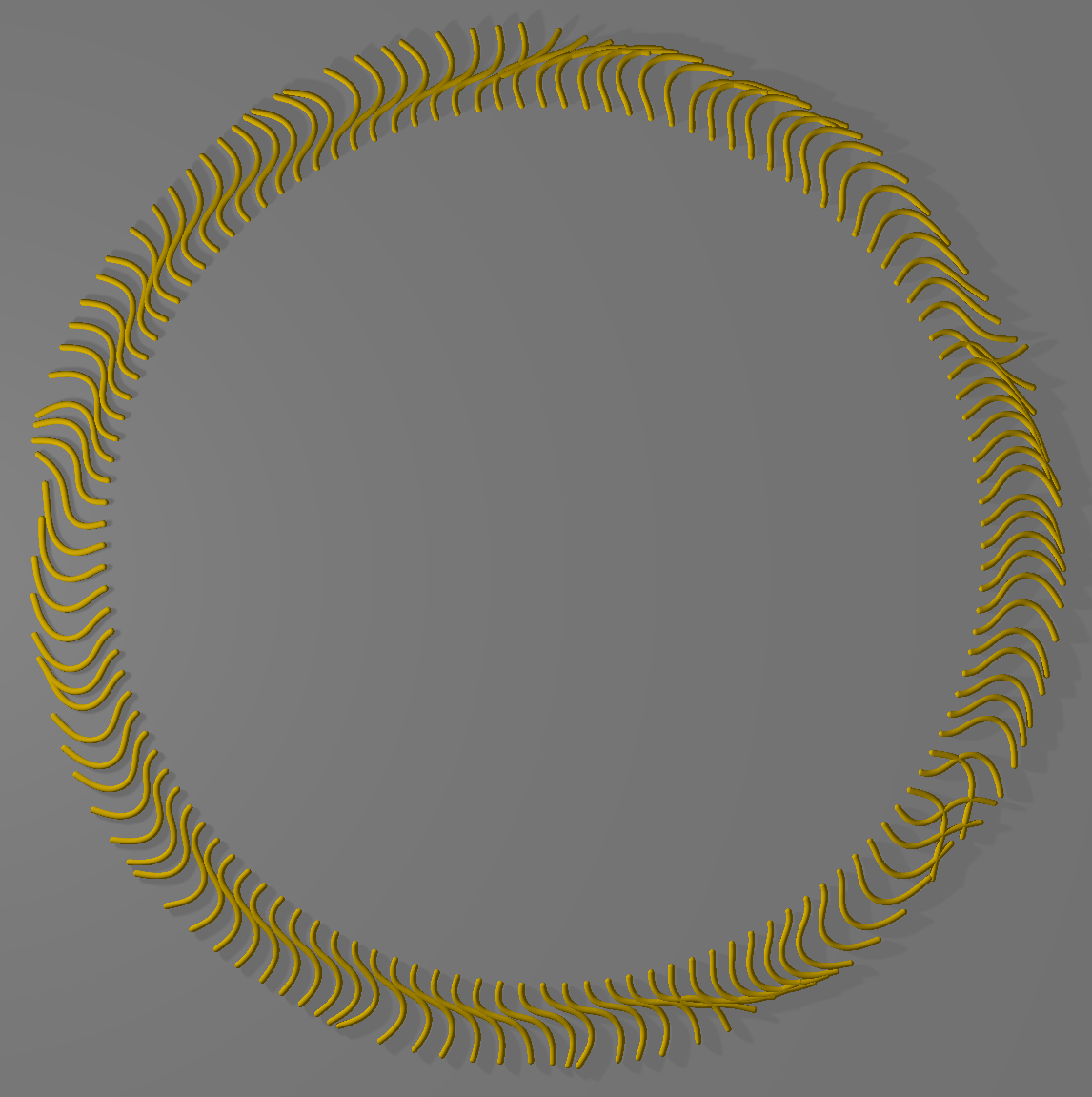}
}
\qquad
\subfloat[][]{
\includegraphics[width =\columnwidth]{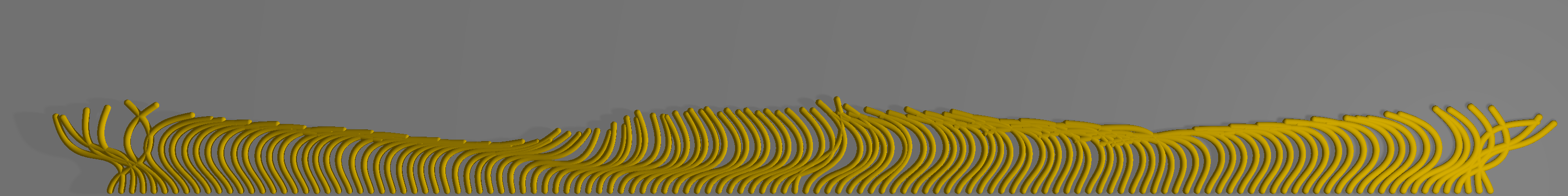}
}
\caption{Simulated metachronal wave formation for 128-polymer arrays in (a) circular and (b) planar geometries. 
In both cases, $k_{oseen} = 0.1$, $k_{stiff}  = 10.0$, $H=1$.
\label{fig:anims}}
\end{figure}
Videos of select simulations are included in the Supplementary
Materials. Fig. \ref{fig:anims} shows some still frames of simulated
arrays demonstrating metachronal wave behavior in both the planar
and circular geometries.

\begin{figure*}
\centering
\subfloat[][]{
\includegraphics[width=0.15\textwidth]{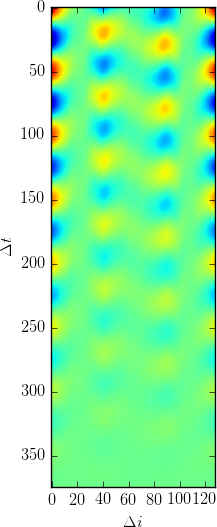}
}
\subfloat[][]{
\includegraphics[width=0.15\textwidth]{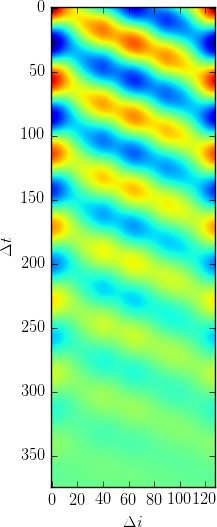}
}
\subfloat[][]{
\includegraphics[width=0.15\textwidth]{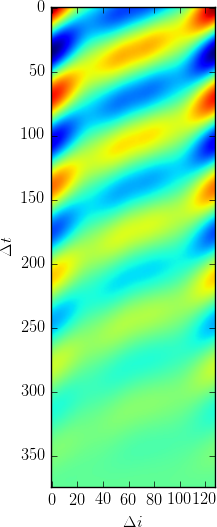}
}
\qquad
\subfloat[][]{
\includegraphics[width=0.45\textwidth]{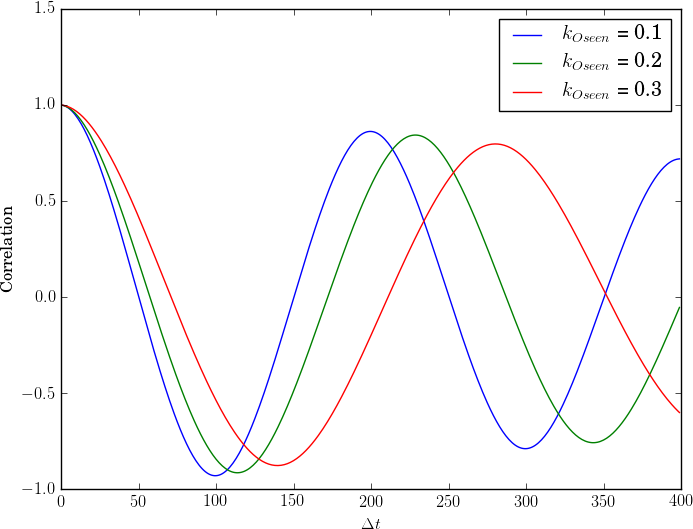}
}
\caption{Full correlation functions for circular geometry with $H=1$
and $k_{stiff}=10$, with $k_{oseen}=$ 0.1, 0.2, and 0.3 (a-c,
respectively). The correlation function at $\Delta i = 0$ for all
of these values of $k_{oseen}$ are shown in (d).  
\label{fig:oseenk}
}
\end{figure*}

\begin{figure*}
\centering
\subfloat[][]{
\includegraphics[width=0.15\textwidth]{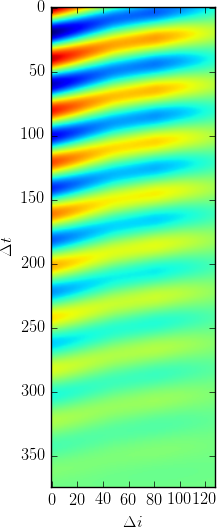}
}
\subfloat[][]{
\includegraphics[width=0.15\textwidth]{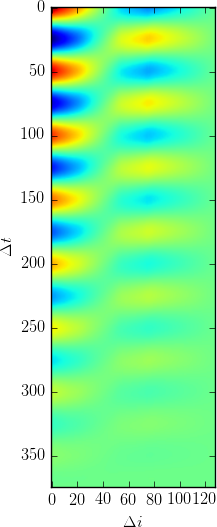}
}
\subfloat[][]{
\includegraphics[width=0.15\textwidth]{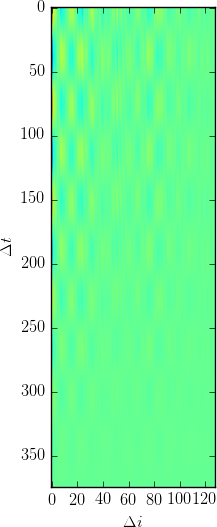}
}
\qquad
\subfloat[][]{
\includegraphics[width=0.45\textwidth]{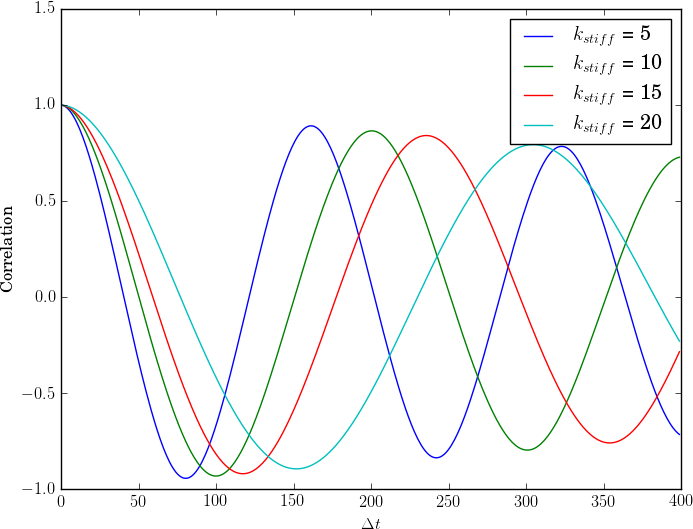}
}
\caption{Full correlation functions for planar geometry with $H=1$
and $k_{oseen}=0.1$, with $k_{stiff}=$ 5.0, 10.0, and 20.0 (a-c,
respectively). The correlation function at $\Delta i = 0$ for all
of these values of $k_{stiff}$ are shown in (d).  
\label{fig:stiffk}
}
\end{figure*}

\begin{figure*}
\centering
\subfloat[][]{
\includegraphics[width=0.15\textwidth]{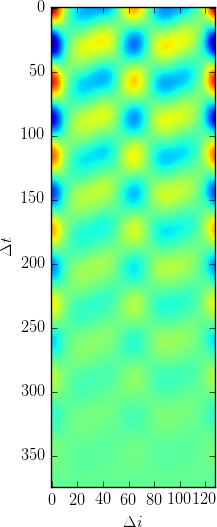}
}
\subfloat[][]{
\includegraphics[width=0.15\textwidth]{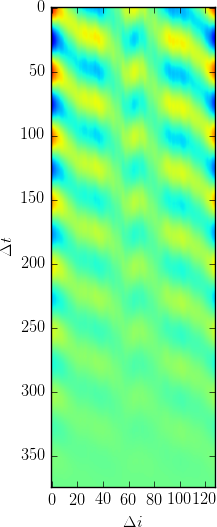}
}
\subfloat[][]{
\includegraphics[width=0.15\textwidth]{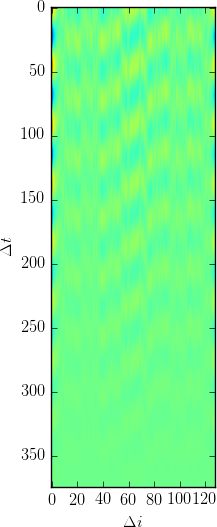}
}
\qquad
\subfloat[][]{
\includegraphics[width=0.45\textwidth]{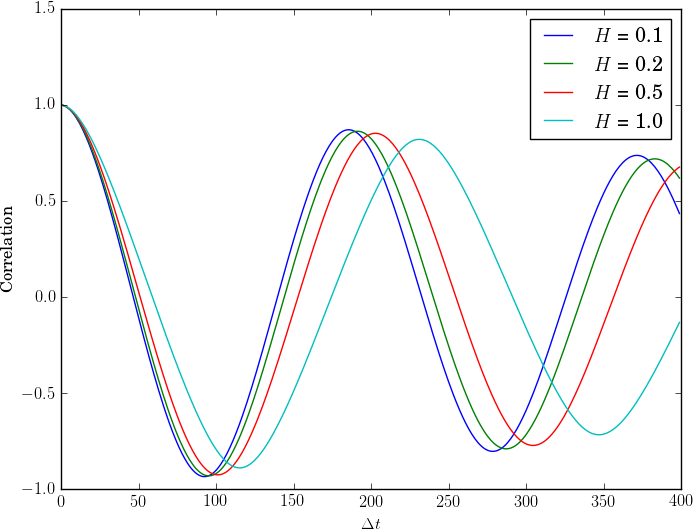}
}
\caption{Full correlation functions for circular geometry with
$k_{stiff}=10$, $k_{oseen}=0.2$, and $H$ = 1.0, 0.5, and 0.1 (a-c,
respectively). The correlation function at $\Delta i = 0$ for all
of these values of $H$ are shown in (d).  
\label{fig:H}} 
\end{figure*}

We characterize the behavior of each system using the correlation function 
for the chain ends $x(i,t)$,
\begin{equation}
C(\Delta i, \Delta t) = \langle \Delta x(i + \Delta i, t+\Delta t)\Delta x(i,t)\rangle,
\end{equation}
where
\[
\Delta x(i,t) = x(i,t) - \langle x(i,t)\rangle.
\]
The average is performed over all chain indices $i$, and time $t$, after a period of equilibration.
Figs.  \ref{fig:oseenk}, \ref{fig:stiffk}, and \ref{fig:H} 
show correlation functions for planar
and a circular geometries (for the circular geometry, the polar angle $\theta$ is
the position variable rather than $x$). 
In the following, we will discuss these and examine how the system responds to changes in
the strength of the interaction tensor, $k_{oseen}$,
$k_{stiff}$, and height $H$. It should be noted that changes in the viscosity or kinesin velocity and density 
(that affect $f_k$), can be absorbed into a rescaling of time, and of $k_{stiff}$.

$k_{oseen}$, has a dramatic
effect on the type of wave behavior seen, or whether it is observed
at all. This strength is a function of the hydrodynamic effects of
kinesin walking along microtubules, and will depend on their density
and speed, as explained in detail in Ref. \cite{Monteith2016}. 
Fig. \ref{fig:oseenk} shows the correlation results of
three 128-polymer simulations in the same circular geometry shown
in Fig. \ref{fig:anims}(a) for three different values of $k_{oseen}$.
There is an overall strengthening of the metachronal
behavior as $k_{oseen}$ is increased from  $0.1$ to $0.2$. 
The sign of the slope reflects the initial conditions of the system.
Long lived waves travel predominantly in a single direction over
long times scales resulting in a 
slope of the crests of the correlation function that can
either be positive or negative. Similar crests are
seen in the analysis of the real experimental data~\cite{Sanchez2011}.
With this circular geometry, the correlation function must be periodic,
which is why it rises again when $i$ becomes large.

The polymer stiffness $k_{stiff}$ also has an interesting effect
on metachronal wave formation. Fig. \ref{fig:stiffk} shows the
correlation functions for $k_{stiff}=$ 5.0, 10.0, and 20.0 in a
planar geometry. While Figs. \ref{fig:stiffk}(a-b) are qualitatively
similar, we do see an apparent decrease in the metachronal wavelength.
Figs. \ref{fig:stiffk}(c-d) show that if the polymer is made too
stiff, no metachronal behavior is observed at all.
In general, planar geometry appears to cause more coherence in the motion of the different
bundles, and the correlation function is dominated by motion at
the longest lengths and time scales. 

The distance between plates, $H$, has a considerable effect on the dynamics
as well. Longer range, more coherent motion is observed when $H$ is larger, and
short range, less coherent motion when $H$ is small. See Fig. \ref{fig:H}. This
is to be expected due to the strong screening effect that these boundary conditions
impose. Smaller $H$ reduces the hydrodynamic coupling, causing a decrease in coherence.

When comparing these results to those of
Sanchez et al., we find that the basic features agree. The videos
included in the supplemental materials qualitatively mimic the
experimental videos, and the experimental correlation analysis
agrees quite well with the simulations. More importantly, this agreement between
theory and experiment was reached from first principles. We only
use a handful of forces in our simulations, and each force has a
physical justification for being used.

There are potential shortcomings of this model that may result in
some differences between experiment and theory. 
The first is that the experiments observe bundles of microtubules
that taper away from their base. The hydrodynamics are not
expected to be uniform along the length of a chain. In addition
these bundles will, for short enough times, behave like rigid
material, but for longer times, because they are connected through
walking kinesin molecules, will behave more as individual mictrotubules
with a greatly reduced elastic constant. On the time scales of the motion,
we expect to be in the latter regime. However the details of the
hydrodynamics and elasticity in these bundles is still not understood
experimentally. In fact, as we mentioned earlier, the polarity of
individual MT's is not known experimentally, and arguments for their
unipolarity are given in the supplementary information, 
S-IV.
But still, the basic mechanism of dynamic buckling due to kinesin drag, 
and metachronal waves being generated by hydrodynamic coupling, is
robust over a wide parameter range, so we believe that these
complications, aside from unipolarity, will not alter the basics of our explanation.

At a more technical level, there are other things that may
make a slight difference to the results here.
The bundles are constrained to move only in the $xy$-plane, and
while it is true that MT motion is nearly 2-dimensional, there is
some room in the $z-$direction that MT bundles can occupy. Additionally,
this model does not account for the fluid boundary condition at the
anchoring surface. This may introduce some errors if a monomer
becomes close ($\sim H$) to the anchoring plane. However, because
of the screening effects of the plates, this should not alter the behavior
at distances large compared to the plate separation. We have tested
for this by adding image charges to the planar case, 
and found that their effects on correlations are small, as expected.

In conclusion,
we have developed a model for the spontaneous formation of wavelike
behavior in active polymer arrays that only requires two ingredients:
semi-flexible chains tethered to a surface, and motors walking from
their bases to their tips. The hydrodynamics in their confined
geometry gives rise to metachronal waves that appear remarkably
similar to what is observed experimentally~\cite{Sanchez2011,sanchez2013engineering}.
There is no need to posit additional mechanisms that force
individual bundles to oscillate. This all happens as a consequence of Newton's
laws and fluid mechanics, allowing us to
gain a better understanding of how metachronal waves form with
considerable predictive power. As such, we have examined new parameter spaces and have
demonstrated boundaries between different types of metachronal
behavior and regimes in which no metachronal behavior exists.
It would be of great interest to test these predictions
experimentally.
Given the simplicity and robust nature of this mechanism, 
and the ubiquity of microtubules and kinesin in cells, it
gives one further impetus to look for other places in biology
where this kind of behavior can be found.

J.M.D. thanks Bill Saxton, Itamar Kolvin, Alex Tayar, and Zvonimir Dogic for useful discussions.
S.E.M. was partially supported by the ARCS Foundation. 
This work was also supported by the Foundational Questions Institute \url{<http://fqxi.org>}.


\bibliography{MT_paper}

\end{document}


\title{Supplementary materials to: Emergence of Metachronal Waves in Active Microtubule Arrays} 
\author{Stephen E Martin}
\affiliation{Department of Physics, University of California, Santa Cruz, CA 95064, USA}
\author{Matthew E Brunner}
\affiliation{Voltaic Inc. 2150 Shattuck Ave, \#704 Berkeley, CA 94704}
\author{Joshua M Deutsch}
\email{josh@ucsc.edu}
\affiliation{Department of Physics, University of California, Santa Cruz, CA 95064, USA}
\date{\today}
\begin{abstract}
\end{abstract}

\maketitle


\section{Supplementary video legends}

\noindent
Videos at \url{https://sites.google.com/ucsc.edu/joshdeutsch/metachronal-videos?authuser=0}

\vskip 0.2in
\noindent
{\bf Supplementary video S1:} 128 microtubule bundles (length 16) with kinesin walkers in a circular
geometry in a fluid chamber with $k_{oseen}=0.1$, chamber height $H=1.0$, $k_{stiff}=10$.
\vskip 0.2in

\noindent
{\bf Supplementary video S2:} 128 microtubule bundles (length 16) with kinesin walkers in a circular
geometry in a fluid chamber with $k_{oseen}=0.2$, chamber height $H=0.1$, $k_{stiff}=10$.
\vskip 0.2in

\noindent
{\bf Supplementary video S3:} 128 microtubule bundles (length 16) with kinesin walkers in a circular
geometry in a fluid chamber with $k_{oseen}=0.2$, chamber height $H=0.2$, $k_{stiff}=10$.
\vskip 0.2in

\noindent
{\bf Supplementary video S4:} 128 microtubule bundles (length 16) with kinesin walkers in a circular
geometry in a fluid chamber with $k_{oseen}=0.2$, chamber height $H=0.5$, $k_{stiff}=10$.
\vskip 0.2in

\noindent
{\bf Supplementary video S5:} 128 microtubule bundles (length 16) with kinesin walkers in a circular
geometry in a fluid chamber with $k_{oseen}=0.2$, chamber height $H=1.0$, $k_{stiff}=10$.
\vskip 0.2in

\noindent
{\bf Supplementary video S6:} 128 microtubule bundles (length 16) with kinesin walkers in a circular
geometry in a fluid chamber with $k_{oseen}=0.3$, chamber height $H=1.0$, $k_{stiff}=10$.
\vskip 0.2in

\noindent
{\bf Supplementary video S7:} 128 microtubule bundles (length 16) with kinesin walkers in a planar
geometry in a fluid chamber with $k_{oseen}=0.1$, chamber height $H=1.0$, $k_{stiff}=5$.
\vskip 0.2in

\noindent
{\bf Supplementary video S8:} 128 microtubule bundles (length 16) with kinesin walkers in a planar
geometry in a fluid chamber with $k_{oseen}=0.1$, chamber height $H=1.0$, $k_{stiff}=10$.
\vskip 0.2in

\noindent
{\bf Supplementary video S9:} 128 microtubule bundles (length 16) with kinesin walkers in a planar
geometry in a fluid chamber with $k_{oseen}=0.1$, chamber height $H=1.0$, $k_{stiff}=15$.
\vskip 0.2in

\noindent
{\bf Supplementary video S10:} 128 microtubule bundles (length 16) with kinesin walkers in a planar
geometry in a fluid chamber with $k_{oseen}=0.1$, chamber height $H=1.0$, $k_{stiff}=20$.
\vskip 0.2in

\noindent
{\bf Supplementary video S11:} 2 microtubules of opposite polarities, green MT's have minus ends on
surface, and blue MT's have plus ends on the surface. There are fixed boundary conditions on the
surface. This shows a simulation for a set of parameters where the two microtubules move. This
behavior was never found when there were more than 2 microtubules in a bundle.
\vskip 0.2in

\noindent
{\bf Supplementary video S12:} Pillar of 9 microtubules of opposite polarities, green MT's have minus ends
on surface, and blue MT's have plus ends on the surface. There are sliding boundary conditions on
the surface. This shows a simulation in a regime with sufficiently weak attractive interactions,
$f_a=1$, where there is a twisting motion inside the pillar but then the minus microtubules suddenly
slide off of the plus ones, finally lying close to parallel with the plane of attachment.

\newpage

\section{The Quasi-2D Interaction Tensor}
\label{app:2Dint}
\begin{figure*}
\begin{center}
\includegraphics[scale=0.6]{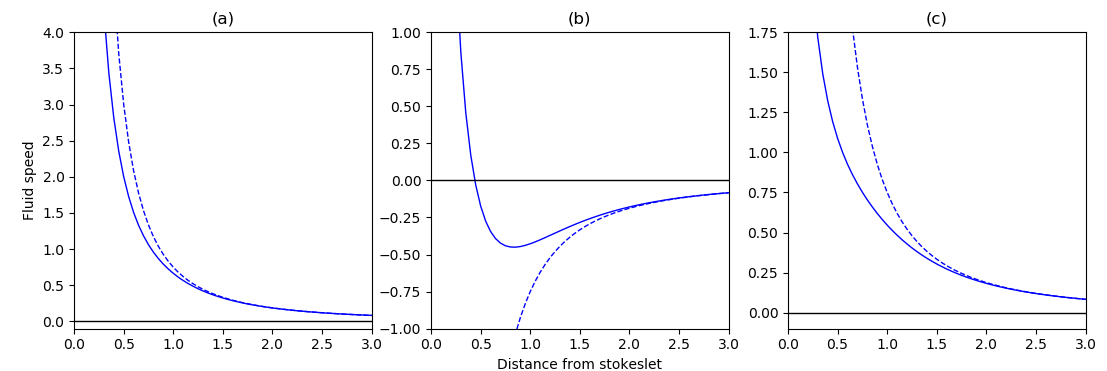}
\caption{Fluid speeds as a function of distance $\rho$ from the stokeslet $\mathbf{F} = \hat \imath$. Solid curves are calculated using the full interaction tensor (\ref{eq:uab}) and dashed lines are the far-field approximation (\ref{eq:uff}). (a) $u_x(\rho)$ along the line $y=0$; (b) $u_x(\rho)$ along the line $x=0$; (c) $u_y(\rho)$ along the line $y=x$. \label{fig:oseen}}
\end{center}
\end{figure*}
\begin{figure}
\begin{center}
\includegraphics[scale=0.35]{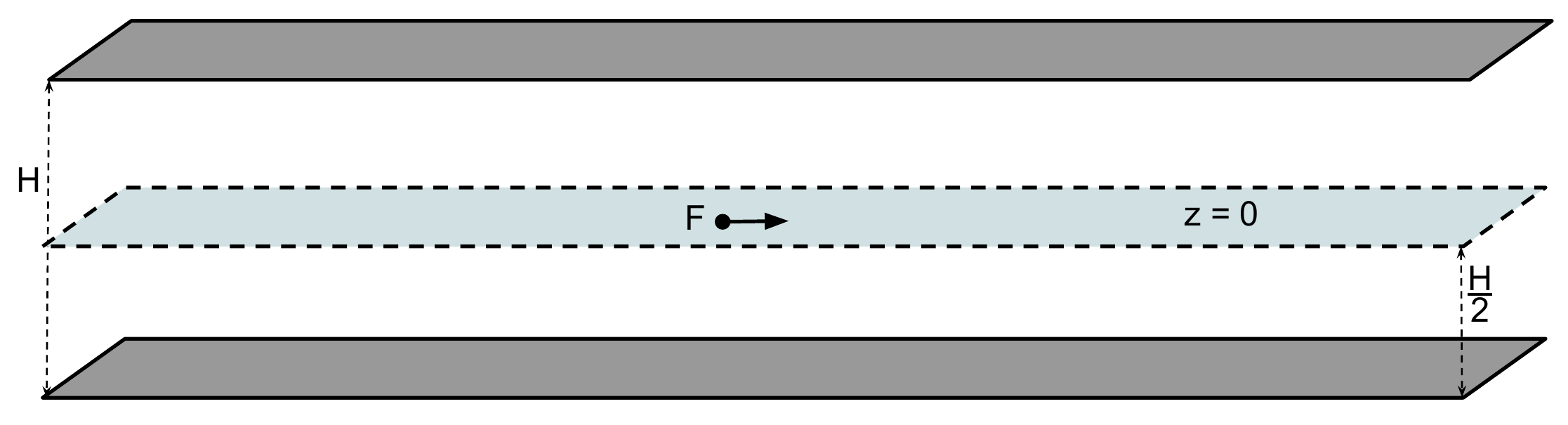}
\caption{Illustration of the geometry for which interaction tensor
is derived in \ref{app:2Dint}. While this is a three-dimensional
system, we constrain polymers to the $xy$-plane. 
\label{fig:schem}}
\end{center}
\end{figure}

The interaction tensor used in simulations is that of a stokeslet enclosed by two infinite parallel plates, as derived by Liron and Mochon\cite{Liron1976}. In general, the interaction tensor $\mathbb{G}$ is defined as the relationship between the fluid flow $\mathbf{u(r)}$ and the stokeslet $\mathbf{F}$ which causes this flow:
\begin{equation}
\label{eq:oseen}
\mathbf{u(r)} = \mathbf{F}\cdot\mathbb{G}(\mathbf{r})
\end{equation}
We assume the system is embedded in a viscous fluid with viscosity $\mu$.
For computational efficiency, we assume all monomers to be only in
the $xy$-plane, with parallel plates at $z = \pm H/2$. This reduces
a three-dimensional problem to two dimensions, as (a) the stokeslet
is located in the $xy$-plane, (b) the stokeslet's direction has no
$z$-component, and (c) we only concern ourselves with flows in the
$xy$-plane (see Fig. \ref{fig:schem}). For
this arrangement, it can be shown from Liron and Mochon's general
result that the interaction tensor a displacement $\mathbf{r}$ (and
$\rho\equiv|\mathbf{r}|$) from a single stokeslet $\mathbf{F}$ at
the origin reduces to
\begin{equation}
\label{eq:uab}
\mathbb{G}(\mathbf{r}) = \frac{H}{8\pi\mu\rho^2}\left\{\left[4\left(\frac\rho H\right)^2 S_1 - \frac12\frac\rho H I_1\right]\mathbb{I}+\left[4\pi\left(\frac{\rho}{H}\right)^3 S_2 + \frac{1}{2}\frac\rho H I_1 -\frac14\left(\frac\rho H\right)^2 I_2\right]\frac{\mathbf{r\otimes r}}{\rho^2}\right\}
\end{equation}
where
\begin{align*}
S_1\equiv&\frac14\sum_{n=0}^\infty \frac{(-1)^n}{\left[\left(\frac{\rho}{H}\right)^2 + n^2\right]^{1/2}}\\
S_2\equiv&\frac{1}{4\pi}\frac\rho H\sum_{n=0}^\infty \frac{(-1)^n}{\left[\left(\frac{\rho}{H}\right)^2 + n^2\right]^{3/2}}\\
I_1\equiv&\int_0^\infty \xi J_1\left(\frac\rho H \xi\right)\frac{\tanh^2\frac\xi 2}{\sinh\xi - \xi}d\xi\\
I_2\equiv&\int_0^\infty \xi^2\left[J_0\left(\frac\rho H \xi\right) -J_2\left(\frac\rho H \xi\right)\right]\frac{\tanh^2\frac\xi 2}{\sinh\xi - \xi}d\xi
\end{align*}
Here, $J_n$ is the Bessel function of the first kind. Because $S_1$ and $S_2$ do not converge rapidly as defined above, we also make use of the Poisson sums
\begin{align*}
S_1 =& \sum_{k=0}^\infty K_0\left[\pi(2k+1)\frac\rho H\right]\\
S_2 =& \sum_{k=0}^\infty (2k+1) K_1\left[\pi(2k+1)\frac\rho H\right]
\end{align*}
where $K_n$ is the modified Bessel function of the second kind.

In the far field, it can be shown that (\ref{eq:uab}) approaches
\begin{equation}
\label{eq:uff}
\mathbb{G}\mathbf{(r)} \approx -\frac{3H}{32\pi\mu\rho^2}\left(\mathbb{I} -2\frac{\mathbf{r\otimes r}}{\rho^2}\right)
\end{equation}
Fig. \ref{fig:oseen} shows plots of $\mathbf{u(r)}$ at selected locations, and compares the exact value from (\ref{eq:uab}) to the far-field approximation from (\ref{eq:uff}).

We can now make some conceptual observations regarding this interaction tensor and how it compares to the boundary-free Oseen tensor $\mathbb{G}_0$:
\[
\mathbb{G}_{0}(\mathbf{r}) = \frac{1}{8\pi\mu r}\left(\mathbb{I} + \frac{\mathbf{r\otimes r}}{r^2}\right)
\]
First, we immediately notice a $1/r$ dependence (rather than
$1/\rho^2$). This means forces without boundaries tend to be more
long-range, and boundaries result in long-range screening. Second,
$\mathbb{G}_{0}$ is always positive, whereas this is not true for
the interaction tensor used here. One key implication of this is
that flows created by a stokeslet are often flowing opposite its
direction (e.g. Fig. \ref{fig:oseen}b). Both of these qualities
may enhance metachronal behavior in the confined system. Screening
means that interactions between nearby polymers are most important,
creating a ``domino effect" from one polymer to the next rather
than having motion more influenced by long-range interactions. The
creation of opposing flows means (among other things) that if one
polymer is moving toward the anchoring surface, it may exert a force
on many of its neighboring polymers \textit{away} from the anchoring
surface. This encourages wavelike behavior rather than uniformity
of beating motion.

\section{Simulation Methods}
\label{app:sim}

\begin{figure}
\begin{center}
\includegraphics[width = 3in]{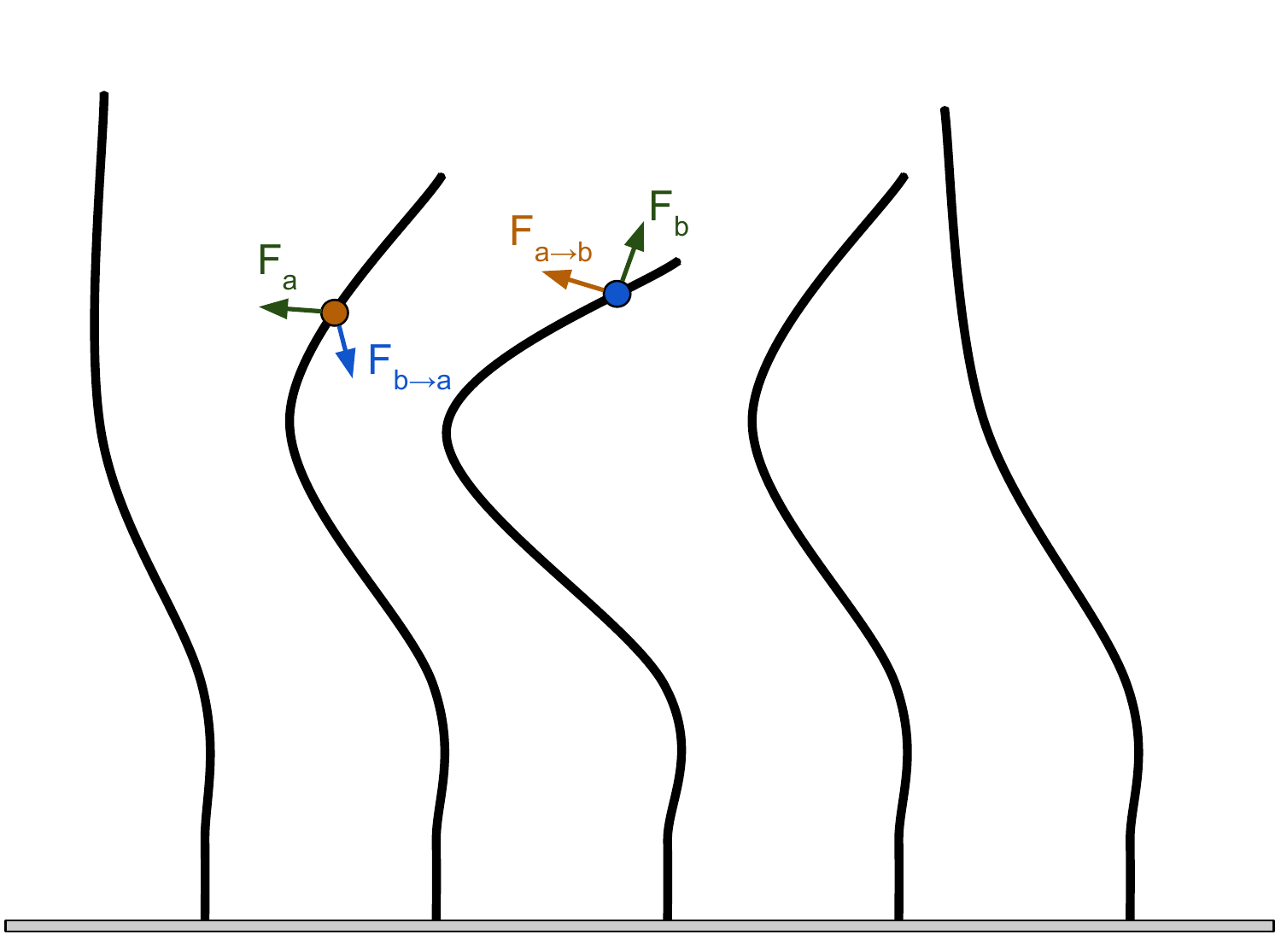}
\caption{Illustration of hydrodynamic forces between two example
monomers in a planar polymer array. The green forces are the sum
of non-hydrodynamic forces on the monomer (and by extension the
force the monomer exerts on the surrounding fluid). $\vec F_{a\rightarrow
b}$ and $\vec F_{b\rightarrow a}$ are the hydrodynamic forces on
monomer $b$ due to $\vec F_a$ and the hydrodynamic force on monomer
$a$ due to $\vec F_b$, respectively. 
\label{fig:hydroforces}}
\end{center}
\end{figure}

The algorithm we implement is built on work that was used to
simulate the mechanism behind cytoplasmic streaming in \textit{Drosophila}
oocytes \cite{Monteith2016}, and many of the methods and equations
below are explained in detail in these papers. This software simulates
an array of active microtubules tethered to a plane that works as follows
and is explained in further detail below.
\begin{enumerate}
\item After an array of polymers is initialized, forces on all
monomers are summed (described below, also see Fig.
1
of main paper) and monomer position and
velocity are updated using time step $dt$.
\item This motion initiates complex flow in the surrounding fluid.
The fluid flow is not simulated directly, but the resulting
hydrodynamic forces from this flow are calculated via an Oseen
tensor with corrections by Blake\cite{Blake1971}. This is illustrated in Fig. \ref{fig:hydroforces}.
\item Forces on each monomer are summed, and monomer position and
velocity are updated accordingly.
\item Once updated, steps 2-3 are repeated.
\end{enumerate}
In the present work there were these differences: 
\begin{enumerate}
\item $N$ Microtubules are confined to the $xy$-plane, with polymer
bases separated by a distance $l$ tethered either to a flat plate
at $y=0$ or to a circular boundary. For all presented results,
$N=128$. The geometry of this is shown in Fig. \ref{fig:schem}.
\item At the tethering point, a potential was added in order to
keep the base monomer approximately orthogonal to the boundary.
\item Rather than the Blake correction to the Oseen tensor, we use
the simplified Liron/Mochon interaction tensor described in supplemental Section \ref{app:2Dint}.
We also investigated varying the  hydrodynamic coupling parameter $k_{oseen} \equiv 1/(8\pi\mu)$.
\end{enumerate}

\noindent
Now we describe how the above was accomplished in more detail.
Each polymer is composed of $n=16$ monomers. The $i$th monomer
position $\mathbf{r}_i$ is updated a using a fourth order Runge
Kutte integration of the equation
\begin{equation}
\frac{d\mathbf{r}_i}{dt} = \mathbf{u(r}_i) - k_{kin}\left(\textbf{r}_{i-1} - \textbf{r}_{i+1}\right)
\end{equation}
where $dt$ is the time step (set to 0.003), $k_{kin}$ (set to 0.2)
controls the strength of the kinesin force tangent to the polymer
($\mathbf{F}_{kin}$ in Fig. 
1),
and $\mathbf{u(r}_i)$
is the fluid velocity due to the motion of all other monomers as
given by Equation \ref{eq:oseen} and \ref{eq:uab} (which imparts
the forces $\mathbf{F}_{a\rightarrow b}$ in Fig. \ref{fig:hydroforces}):
\begin{equation}
\mathbf{u(r}_i) = \sum_{j\neq i} \mathbf{F}_j \cdot \mathbb{G}(\mathbf{r}_i-\mathbf{r}_j)
\end{equation}
Here, $\mathbf{F}_j$ is the total force on the fluid due to the $j$th monomer. Because there are are no inertial effects when $Re\ll 1$, any non-hydrodynamic force exerted on the monomer must be transferred to the fluid. In our case,
\begin{equation}
\mathbf{F}_j = \mathbf{T}_j + \mathbf{C}_j + \mathbf{Q}_j,
\end{equation} 
where
\begin{itemize}
\item $\mathbf{T}_j = k_{spr}\left[\left(|\mathbf{r}_{j-}|-\ell\right)\hat{\mathbf{r}}_{j-} + \left(|\mathbf{r}_{j+}|-\ell\right)\hat{\mathbf{r}}_{j+} \right]$\\
with $\mathbf{r}_{j\pm}\equiv \mathbf{r}_{j\pm 1} - \mathbf{r}_j$, is the spring force keeping monomer separation approximately constant. For our simulations, $k_{spr}=100$ and $\ell=1$.
In these simulations the separation between polymer bases defines above, $l$ is equal to $4\ell$.
\item $\mathbf{C}_j = k_{stiff}\left(2\textbf{r}_i - \textbf{r}_{i+2} - \textbf{r}_{i-2}\right)$\\
is the stiffness force which resists polymer bending. $k_{stiff}$ is varied in our simulations, but typically $5 \leq k_{stiff} \leq 20$.
\item $\mathbf{Q}_j = \mathbf{P}_j + \mathbf{B}_j + \mathbf{W}_j + \sum_k\mathbf{H}_{jk}$\\
is the sum of miscellaneous conditional forces:
\begin{itemize}
\item $\mathbf{P}_j = k_{pin}\left(\mathbf{r}_j - h\boldsymbol{\hat{\jmath}}\right)$\\
\phantom{.}\hfill if $(j \mod n) = 1$\\
is the force on the base monomer of each polymer chain keeping it pinned to the anchoring surface. For our simulations, we set $k_{pin} = 100$ and $h=1$.
\item $\mathbf{B}_j = k_{pin2}\left(\mathbf{r}_j - \mathbf{r}_{j-1} - \ell\boldsymbol{\hat{\jmath}}\right)$\\
\phantom{.}\hfill if $(j \mod n) = 2$\\
is the force on the second monomer in each polymer chain, keeping
the base of each polymer approximately orthogonal to the anchoring
surface ($F_{base}$ in Fig. 
1).
For our simulations,
we set $k_{base}=100$.
\item $\mathbf{W}_{j} = k_{wall}\left[1-\left(\frac{d_{wall}}{y_j}\right)^4\right]\boldsymbol{\hat\jmath}$\\
\phantom{.}\hfill if $y_j < d_{wall}$\\
is the repulsive force exerted by the anchoring plane on any monomer that gets close to the wall. For our simulations, we set $d_{wall}=0.5$ and $k_{wall} = 100$.
\item $\mathbf{H}_{jk} = k_{rep}\left[1-\left(\frac{d_{rep}}{|\mathbf{r}_j - \mathbf{r}_k|}\right)^4\right]\left(\mathbf{r}_j - \mathbf{r}_k\right)$\\
\phantom{.}\hfill if $|\mathbf{r}_j-\mathbf{r}_k| < d_{rep}$\\
is the repulsive force between monomers that are very close to one another. For our simulations, we set $d_{rep}=0.5$ and $k_{rep} = 1$.
\end{itemize}
\end{itemize}
\section{Analysis of Unipolarity}
\label{app:unipolar}

The work of Sanchez et al.~\cite{Sanchez2011,sanchez2013engineering} consists of a mixture of 
biotin-labeled kinesin-1 motors bound together to form clusters using multimeric streptavidin
and taxol stabilized microtubules in a polyethylene-glycol solution with
ATP. These form bundles of microtubules, some of which are adsorbed to
air-water or air-glass interfaces, that point out from the interface forming a
lawn of microtubule bundles. These bundles are flexible and show bending similar
to what is seen in the simulations described here
in both the time scales, length scales, and correlations
between different bundles.

The question that is not answered in the experimental work is the directionality
of the microtubules inside a bundle. The microtubules forced into bundles by the polyethylene glycol (PEG)
could be of mixed polarity
so that some have their minus ends at the interface while others have their plus
ends there. We will refer to microtubules with different orientations as having
different ``polarities", minus-ends against the interface as ``minus" and those with opposite
polarity as ``plus". 

The problems with having a mixed polarity bundle are two fold. The first is
that for a wide range of experimental parameters, we expect mixed polarity 
bundles to be unstable~\cite{kruse2000actively,liverpool2003instabilities}. The second problem is that it is not clear that
mixed polarity bundles can give rise to the motion seen experimentally. We
will analyze both problems below.

\subsection{Instability of mixed polarity bundles}

The first problem is that adjacent
microtubules with different polarities will be linked by kinesin clusters that
will apply equal and opposite forces to them. This will cause the minus
microtubules to be pushed toward the interface, and the plus ones away from it. 
The forces from the kinesin act in parallel on a microtubule over its length
which is of order $10 \mu m$.
The forces that these cause can be competitive with depletion forces caused by the
PEG as we will now see.
A full analysis of this is not possible without more information
about the details of the system such as the density of kinesin clusters and chain
lengths of the PEG. However we can do a calculation to show that even with
very modest assumptions concerning kinesin density, expulsion of plus
microtubules will take place.

Depletion forces exert an osmotic pressure on microtubules and filaments. 
Each polymer excludes a roughly spherical region of order its radius of
gyration $R_g$.
Entropic forces favor the separation of microtubules into bundles because less
volume is excluded by the PEG. We will estimate the force acting on a
single microtubule protruding from a bundle. 
PEG is depleted in a region of size $R_g$ around the microtubule. The increase in
free energy per unit area caused by this depletion is of order $p R_g$  where the osmotic pressure is  $p = k_B T \rho$, and
$\rho$ is the number of polymers per unit volume. The increase in free energy
$dF$, in raising the microtubule by a height $dz$, is $dF = (2\pi R_m dz) p x$. Here $R_m$
is the microtubule radius. If we assume that the
polymers are close-packed around the microtubule to get the maximum effect, then $\rho = 1/(4 \pi R_g^3/3)$.
So the force needed to push the microtubule out of the tip of the bundle is
$f = dF/dz = (3/2) R_m k_B T/R_g^2$.

$R_m \approx 13 nm$ and conservatively taking $R_g = 1nm$, which is quite small
for PEG, $f = 81 pN$
The stall force of kinesin is approximately $5pN$~\cite{meyhofer1995force}. So only 16.2 kinesins are needed to overcome
the depletion forces and expel this microtubule from the bundle. 

The minimum separation of kinesin on a microtubule is $8nm$ and there are 13
tracks around its circumference. Because kinesin has a strong affinity for
microtubules we expect a high density of bound kinesin. Therefore 16 kinesins
contributing to the force over
a distance of $10 \mu m$ is over three orders of magnitude less dense
than the maximum density attainable.
This suggests that for a wide range of parameters, the microtubule bundles will
become unipolar with minus-ends against the interface.

\subsection{Model of mixed polarity bundles}

The second problem is
that it is not clear that a mixed polarity bundle can give rise to the motion
seen in experiment. Here we analyze this possibility by using simulation methods
similar to what was used previously to understand molecular motor dynamics~\cite{deutsch2015photomechanical}

We assume that the microtubules are inextensible and that opposite polarity
microtubules apply forces in equal and opposite directions. We discuss the
different forces separately.

First there is an effective attractive interaction between microtubules independent of 
their polarities induced by the presence of PEG polymers. We choose a short range force so the monomers separated by a
distance $\br$  within a range $\sigma_s$ will feel an attractive force due to
depletion forces as discussed above. To
simplify the expressions we use a normalized unitless distance $\Delta \equiv
\br/\sigma_s$.
The force between any two monomers for $\Delta < 1$ is taken to be
\begin{equation}
{\bf f}_{attr} = f_a \Delta^4 (1-\Delta^{12})^3 \br
\end{equation}
where $f_a$ is the strength of the attractive interaction.
The reason for choosing this functional dependence on $\Delta$ was to produce
a force that was close to constant for $\Delta < 0.6$, and then drop
smoothly to zero, so as to work well with the Runge Kutte algorithm.

Second, we introduce an even shorter range repulsion between monomers
that diverges at a hard core radius $\sigma_h$ and goes to zero at $\sigma_s$:
\begin{equation}
{\bf f}_{rep} = f_r\left(\frac{1}{r^2-\sigma_h^2}- \frac{1}{\sigma_s^2-\sigma_h^2}\right)^4 \br 
\end{equation}
where $f_r$ is the strength of the repulsive interaction.

Third, we introduce an equal and opposite forces between monomers on opposite polarity microtubules
that are within a distance $\sigma_s$.
The direction of the force is as follows. We compute the tangents to both
monomers as $(\br_{i+1}-\br_{i-1})/2$. Then we choose the direction $\bf t$, to be the
average of these two tangents. The magnitude of the kinesin force is
\begin{equation}
{\bf f}_{kin} = f_k (1-\Delta^{12})^3 
\end{equation}
where $f_k$ is similar the symbol used previously in the main text and denotes the magnitude of
the kinesin force.

These forces are added to the elastic forces, viscous drag, and tension that
must be introduced to conserve link length and the equation of motion is iterated
using a method for updating chains with constant link length~\cite{deutsch1988theoretical,deutsch1989theoretical}.

We also tried two separate kinds of boundary conditions. First, tethering the chains to
fixed points on the surface which we will call ``fixed" boundary conditions. Second, confining the chain ends to a two
dimensional plane but letting the ends move within that plane, which we will
call ``sliding" boundary conditions. 

We tried a wide range of parameters, of different elastic constants, attractive
interactions, number of microtubules, and boundary conditions. What we found is
now summarized. 

For two chain bundles of opposite polarity we did find a set of parameters which
showed movement of the bundle with: 
$f_r =  10.0$, $\sigma_s =  2$, $\sigma_h = 1$, $f_a =  3$, $f_k = 0.2$, $k_{stiff}=
100$, and chain length of $20$, see supplemental move S11.

For larger bundle sizes, e.g. 9 chains,  we did not find anything similar to
experiments. With fixed boundary conditions, and started as a pillar of
parallel microtubules with slightly randomized directions, the chains would settle down to
a pillar shape that would not change with time for sufficiently small attractive
interactions $f_a$, but when this became greater than a certain value that
depends on elastic constant and other parameters, it would suddenly collapse
into a ball because this is more highly favored energetically.

When we chose sliding boundary conditions, and for sufficiently weak attractive
interactions, $f_a=1$ there was a regime where there was twisting motion inside the
pillar but then the minus microtubules would suddenly slide off of the plus
ones, finally lying close to parallel with the plane of attachment, see supplemental movie S12. 
It therefore appears that a two microtubule bundle moves because of a strong
anisotropy in forces seen in cross sections. In larger bundles, the forces
through the bundle are more homogeneous which acts to stabilize them.

We conclude that by direct physical modeling of a mixed polarity bundle, it is
not clear if there are any reasonable parameters which show motion similar to
what is seen in the experiments of Sanchez et al~\cite{Sanchez2011,sanchez2013engineering}.

Note that the elastic constant of a microtubule in a bundle will depend strongly on the
rate at which it is bent. For very short times, the bonds between different
microtubules caused by kinesin binding will be fixed in position giving the bundle the
elastic constant of a cylinder of radius $R$ which is $\propto R^4$. However the
oscillations here take place on minute timescales. In that case the individual
kinesin molecules have velocities of order $1\mu m/s$ so 
they unbind and move very far
on this time scale. This allows neighboring microtubules to move relative to
each other, to eliminate stress. Therefore on sufficiently long timescales,
this reduces the elastic constant of a microtubule to that of one in isolation.


\bibliography{MT_paper}